\documentclass[aps,prm,twocolumn,superscriptaddress,showkeys]{revtex4-1}

\usepackage{graphicx}
\usepackage{dcolumn}
\usepackage{bm}

\usepackage{blindtext}

\makeatletter
\def\p@subsection{}
\makeatother

\begin{document}



\title{Classification of Local Chemical Environments from X-ray Absorption Spectra using Supervised Machine Learning}


\author{Matthew R. Carbone}
\affiliation{Department of Chemistry, Columbia University, New York,
New York 10027, USA}
\affiliation{Computational Science Initiative, Brookhaven National Laboratory,
Upton, New York 11973, USA}

\author{Shinjae Yoo}
\affiliation{Computational Science Initiative, Brookhaven National
Laboratory, Upton, New York 11973, USA}

\author{Mehmet Topsakal}
\email[]{mtopsakal@bnl.gov}
\affiliation{Center for Functional Nanomaterials, Brookhaven National
Laboratory, Upton, New York 11973, USA}

\author{Deyu Lu}
\email[]{dlu@bnl.gov}
\affiliation{Center for Functional Nanomaterials, Brookhaven National
Laboratory, Upton, New York 11973, USA}


\date{\today}

\begin{abstract}
X-ray absorption spectroscopy is a premier, element-specific technique
for materials
characterization. Specifically, the x-ray absorption near edge structure
(XANES)
encodes important information about the local chemical environment of an
absorbing atom, including
coordination number, symmetry and oxidation state. Interpreting XANES
spectra is a key step
towards understanding the structural and electronic properties of
materials, and as
such, extracting structural and electronic descriptors
from XANES spectra is akin to solving a challenging inverse problem.
Existing
methods rely on empirical fingerprints, which are often qualitative or
semi-quantitative and not transferable. In this study, we present a machine
learning-based approach, which is capable of classifying the local coordination environments of the absorbing atom from simulated K-edge XANES spectra. The machine learning classifiers can learn important spectral features in a broad energy range without human bias, and once trained, can make predictions on the fly. The robustness and fidelity of the machine learning method are demonstrated by an average $86$\% accuracy across the wide chemical space of oxides in eight $3d$ transition metal families. We found that spectral features beyond the pre-edge region play an important role in the local structure classification problem, especially for the late $3d$ transition metal elements.
\end{abstract}


\maketitle

\section{Introduction \label{intro}}


Knowledge of material structures at the atomic scale is essential to understanding physical phenomena
and material properties that can lead to practical applications.
Specifically, key information about the local chemical environment (LCE)
surrounding an atom, including
symmetry, coordination number, bond length and bond angle,
forms the fundamental basis that determines the electronic properties of materials.
In order to resolve the structure-property relationship, the characterization of atomic structures and their dynamic changes under different thermodynamic conditions has become a primary target
of experimental studies. Such efforts have made tremendous impact on many research
fields, including superconductivity~\cite{Jorgensen1987}, ultrafast dynamics~\cite{Cavalleri2001}, energy storage~\cite{Kang2009}, and photocatalysis~\cite{Peng2015}. Recent progress in materials discovery using smart automation~\cite{Tabor2018,Gubernatis2018} and \emph{in situ} and \emph{operando} experiments~\cite{Meirer2018} further highlights emerging challenges and opportunities of materials characterization in \emph{real time}.

Amongst many experimental techniques (e.g. imaging, diffraction, and spectroscopy), the x-ray absorption
near edge structure (XANES) is a premier tool for probing LCEs, because it is element specific, sensitive
to local structural and electronic properties, and
applicable under harsh experimental conditions~\cite{Ankudinov2002,Rehr2009,Frenkel2012},
making it a robust structure refinement method~\cite{Ankudinov1998,Ankudinov2002,Bazin2003,Ciatto2011,Ma2012,Kuzmin2014}.
Given the atomic arrangement of a sample ($\mathbf{x}$), its XANES spectra
($\mathbf{y}$) can be determined through
quantum mechanical laws ($f$) via the mapping $\mathbf{y}=f(\mathbf{x})$. 
Extracting the information of the LCE ($\tilde{\mathbf{x}}$) as a subset of $\mathbf{x}$ from spectral data can be formulated as an inverse problem:
$\tilde{\mathbf{x}}=f^{-1}(\mathbf{y})$. The solution of this inverse problem is highly nontrivial, because the spectral information in experimental XANES is not only abstract, but also averaged over the whole sample. Consequently, much of the success in the past has been achieved by using fingerprints established from empirical observations.

In this study, we focus on $3d$ transition metal K-edge XANES, which carries rich
information about the electronic transitions from the $1s$ core level of the absorbing atom to unoccupied
states. Since the 1940's, extensive research has been carried out to correlate spectral
features of K-edge XANES spectra, especially in the pre-edge region, to LCEs~\cite{Srivastava1973,Yamamoto2008}. For example, Hanson \emph{et al.}~\cite{Hanson1949} observed
distinct chemical shifts in the absorption edge of Mn K-edge XANES in Mn, MnS, MnO$_2$ and KMnO$_4$.
Wong \emph{et al.}~\cite{Wong1984} showed linear relationships between the oxidation state of V and both pre-edge and absorption edge positions in the K edge. Farges \emph{et al.}~\cite{Farges1996, Farges1997,Farges2001} and Jackson \emph{et al.}~\cite{Jackson2005} conducted comprehensive studies of the correlation
between pre-edge features and the coordination number in Ti, Fe and Ni compounds;
they found that the pre-edge peak intensity decreases with increasing coordination number. For fixed coordination number, early 3$d$ transition metal elements (Ti, V, Cr and Mn) have stronger pre-edge peaks than late transition metal elements (Fe, Co, Ni and Cu) overall~\cite{Yamamoto2008}.
Furthermore, while both pre-edge peak locations and intensities in Ti~\cite{Farges1996} and Ni species~\cite{Farges2001} exhibit a significant dependence on the coordination number, the pre-edge peak positions in Fe compounds are independent of coordination
number~\cite{Jackson2005}.

From a theoretical standpoint, the pre-edge peak intensity can be
understood qualitatively from quantum mechanical selection rules. The dominant contribution in K-edge
XANES comes from $s \rightarrow p$ dipole transitions, as the $s \rightarrow d$ quadrupole terms are
generally orders of magnitude smaller. The density of states corresponding to the pre-edge regions of 3$d$ transition metals are derived primarily from their empty 3$d$ bands, and direct $s \rightarrow d$
transitions are dipole-forbidden, which implies a vanishing peak intensity. However, pre-edge peak intensity is enhanced when atomic, unoccupied $p$ and $d$ states hybridize.
According to group theory, atomic $p-d$ mixing is allowed under T$_d$ symmetry, but is forbidden under O$_h$
symmetry~\cite{Cotton1956A,Cotton1956B}. As a result, 3$d$ transition metals with tetrahedral geometries tend to exhibit
stronger pre-edge peak intensities than those with octahedral geometries. To this end, empirical diagrams have been compiled to classify four-, five-
and six-coordinated Ti, Ni and Fe based on pre-edge peak positions and intensities~\cite{Farges1996, Farges1997,Farges2001,Jackson2005}; we will refer to this method as the empirical fingerprint approach. 

Despite the wide range of applications of the empirical fingerprint approach, including classifying LCEs in crystals, amorphous systems~\cite{Farges1997,Mountjoy1999} and catalysts~\cite{Bordiga2013}, it has several limitations that may hinder its practical applications in the broader materials domain. First, coordination number is not the only factor that affects pre-edge peak features. Quantitative pre-edge features are determined by multiple factors, including coordination number, local distortion, oxidation state, and the nature of the ligands~\cite{Jiang2007}. For example, local distortions, e.g. displacements from the inversion center in octahedral geometries, under the crystal field can lower the local symmetry and enable atomic $p-d$ mixing, resulting in dramatic enhancement of pre-edge peak intensity~\cite{Jiang2007}. Such local distortion-induced pre-edge peak intensity enhancement has been reported in the V K-edges of six-coordinated MgV$_2$O$_6$~\cite{Yoshida1992} and NaV$_{10}$O$_{28}$~\cite{Tanaka1988}, and in the Ti K-edge of six-coordinated Li$_4$Ti$_5$O$_{12}$~\cite{Zhang2017}.
Therefore, isolating pure LCE effects and extracting robust correlations between the LCE and simple spectral descriptors, although valid for exemplary systems, may not be feasible for more structurally complex ones. 

Secondly, the empirical fingerprint approach relies on human knowledge to engineer spectral descriptors, which may introduce bias. For example, existing spectral descriptors are primarily derived from the pre-edge region (e.g. peak positions and intensities). However, it is known that pre-edge features are much less visible in late transition metals than early transition metals~\cite{Yamamoto2008}. Therefore, existing empirical fingerprint approach may not work effectively for late transition metals due to poor spectral contrast in the pre-edge region. One may need to systematically explore main- and post-edge spectral features in order to engineer and optimize new descriptors, which may not necessarily be simple ones, to tackle this problem.

Machine learning (ML) methods are a promising candidate to solve this inverse materials characterization problem. Instead of relying on empirical features derived from a small number of human observations, ML methods are data-driven approaches that make predictions based on large training sets, eliminating human bias from the feature selection process. There are myriad successful examples of the utilization of ML methods in condensed matter physics, materials science and chemistry, including methods to solve many-body problems~\cite{Carleo2017}, predict quantum phase transitions~\cite{Nieuwenburg2017}, generate force field potentials~\cite{Behler2007}, design new catalysts~\cite{Kitchin2018}, and perform structure refinement~\cite{Timoshenko2017,Timoshenko2018}. In the context of XANES, one expects ML algorithms to \emph{learn} spectral descriptors in the full energy range of the spectrum and weight them appropriately for robust LCE predictions.

In this study, we tackled the LCE classification problem using supervised ML applied to a wide energy range of the XANES spectra ($\sim 50$ eV above the onset) as input. In this way, the spectral feature space was systematically explored in order to establish the relationships between XANES spectra and LCE classes, specifically
the local atomic geometries. As proof-of-principle, we applied ML algorithms to synthetic K-edge XANES spectra obtained from high throughput \emph{ab initio} calculations. This study serves as a precursor to a potentially very powerful tool for real time structure refinement using experimental XANES, which will require in-depth understanding of the accuracy of the theory and further improvement of the ML algorithms.



\section{Methods \label{methods}}

\begin{figure}[!htb]
\includegraphics[scale=0.0327]{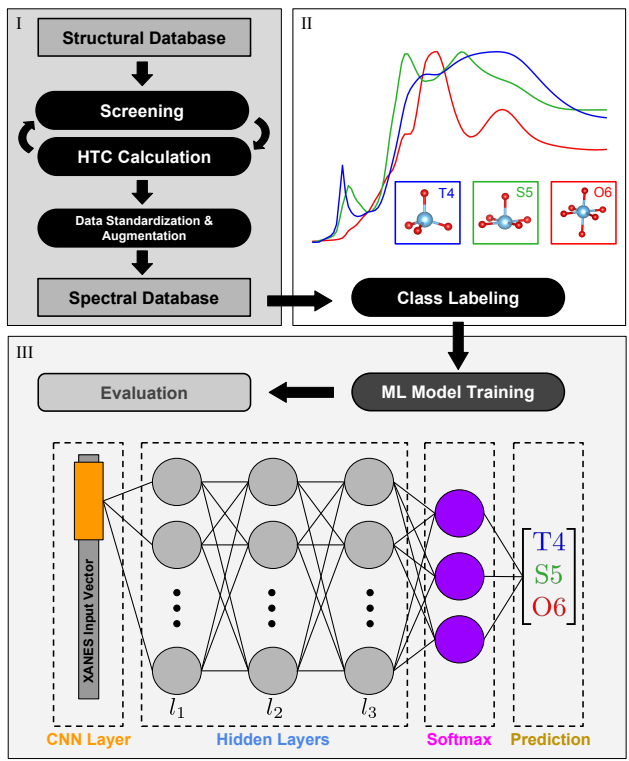}
\caption{\label{fig:QAQC} 
Workflow of the spectrum-based local chemical environment classification framework using supervised machine learning, which contains
three modules: (I) data acquisition supplemented by high throughput computing (HTC) calculations, (II) labeling and (III) training of machine learning models. The machine learning
architecture (set of hyperparameters) used in this work is shown in Module~III.
Notably, the model consists of an optional convolutional layer (shown in orange) followed by
three hidden layers $l_1, l_2$ and $l_3$ consisting of 90, 60 and 20 neurons, respectively,
ending with a softmax output. Further details of
the network are described in Subsection~\ref{training machine learning models}.}
\end{figure}
The workflow of the element-specific, spectrum-based LCE classification framework is
summarized in Figure~\ref{fig:QAQC}, which contains three core modules: \emph{data acquisition}, \emph{LCE class labeling} and \emph{training of machine learning models}. We stress that the workflow we
developed can be adapted 
to a wide range of elements characterized by different spectroscopic techniques, as long as the spectral information
is element specific and sensitive to the LCE such that there exists distinguishable spectral contrast associated with different LCEs. Below, we describe each module in detail.


\subsection{Data Acquisition}
For any given element, the first step is to extract atomic structures representing different LCEs 
from existing materials structure databases. The structural database must be
large enough to build a reasonably-sized training set for machine learning
models. To demonstrate the applicability of the LCE classification framework, we have considered eight
3$d$ transition metal elements
(Ti, V, Cr, Mn, Fe, Co, Ni, and Cu) and extracted all available oxide structures that have been structurally optimized using density functional theory (DFT)
from the Materials Project Database~\cite{Jain2013,Ong2013,Ong2015}.

Once the structural database is established, the next step is to generate the corresponding spectral
database. We focused on the K-edge XANES, as it is element specific and sensitive to the LCE (e.g.,
symmetry, charge state, and coordination number). In principle, one may populate the spectral database
entirely with experimental spectra, but this strategy suffers from several drawbacks. 
First, experimental XANES spectra represent an average of signals
(site-averaged signals) from each absorbing site (site-specific signals). In order to
identify the correlations between spectra and local structures using ML, it is necessary to
use site-specific spectra for training, as they possess much stronger spectral contrast than the
site-averaged spectra. Second, when developing LCE classifiers using supervised ML methods, one
needs to label XANES spectra with
LCE descriptors, which means that only experimental spectra of known structures can be selected for the
database. This requirement severely limits the pool of candidates for the database to mostly well-characterized crystal structures. As a result, qualified experimental spectra
represent only a small fraction of the targeting LCEs in the local configuration space, which are
heavily weighted in known crystals and under-represent the materials space of amorphous systems, surfaces,
interfaces and nanoparticles. Consequently, a pure experimental spectral database suffers from data
availability and data heterogeneity issues.

On the other hand, combined with available structural databases and well-established structure sampling
methods, computational XANES have a clear advantage in exploring the LCE space and producing site-specific
spectra. Furthermore, recent development in computational XANES modeling~\cite{Rehr2000,Taillefumier2002,Gougoussis2009,Prendergast2006,Chen2010, Vinson2011,Vinson2014,Gulans2014,Vorwerk2017,Liang2017,Sun2018} has made it
feasible to contrast experimental spectra quantitatively, enabling accurate local structure refinement of nanoparticles~\cite{Timoshenko2017,Timoshenko2018A}, interfaces~\cite{Small2014}, dopant sites~\cite{Timoshenko2016,Singh2018} and structural phase transformation~\cite{Zhang2017, Timoshenko2018} using computational techniques.
An accurate and computationally efficient first-principles XANES method would be an ideal choice
for sampling the vast LCE parameter space and mitigating data availability and heterogeneity issues. Indeed, computational XANES databases have recently emerged as a new tool for fast structure screening through data mining~\cite{Mathew2018,Zheng2018,Yan2018}.

In this study, we generated the computational XANES database
with the FEFF9 code~\cite{Rehr2010}, which is a popular and computationally efficient method based on
multiple scattering theory. 
We utilized the existing site-specific x-ray absorption spectroscopy FEFF library in the Materials Project~\cite{Mathew2018,Zheng2018} and only calculated spectra not contained in this library. In the first data standardization step, site-specific spectra that failed sanity
checks were automatically discarded, such as when the
FEFF calculations did not converge with the default input or when the output FEFF spectra are not
physical (e.g., with negative absorption coefficients). In the second step, we removed
``duplicates",  which otherwise would have introduced bias if nearly identical structures were selected for both training and testing. We used a site-symmetry finder from the pymatgen library~\cite{Ong2013} to determine
which sites in a crystal structure are symmetrically equivalent. For every pair of spectra, one was
removed from the dataset if the average mean absolute difference between them was less than
$0.015$, a number chosen based on visual inspection of a large number of similar spectra.
The process of removing duplicates also has the benefit of reducing the
total number of necessary FEFF calculations to populate the XANES database. Finally, calculated XANES spectra were spline-interpolated onto an absorbing site-specific energy grid, so that the input feature vector was standardized for each spectrum.
For each type of absorbing site, the energy grid was chosen such that it contains the maximum amount of available information with an energy resolution of approximately 0.5 eV.

All standardized data before augmentation are henceforth referred to as the base
dataset. Training data were augmented by shifting the spectra by $\pm 1$ and $\pm 2$ eV.
The size of the augmented training set thus becomes five times the size of the base dataset. We
found that data augmentation improves the accuracy and robustness of the machine learning model.

\subsection{Local Chemical Environment Class Labeling}\label{feature}

Feature engineering of LCEs is an active research topic with increasing
applications in a variety of areas including, for example, neural network potential development~\cite{Behler2007,Rupp2012,Smith2017,Chimela2017}.
Among many possible choices, labels based on the coordination environment (e.g. tetrahedral, square
pyramidal, and octahedral geometries), although simple, provide key information on chemical bonding and have been widely used in the x-ray spectroscopy community.
In this study, we utilized the continuous symmetry measure (CSM), developed by Avnir and Pinsky~\cite{Pinsky1998} and hosted in the ChemEnv package \cite{Waroquiers2017}, to measure the similarity between an input local geometry and a particular polyhedron. The smaller the CSM for a polyhedron, the more the input geometry resembles it. We applied a cutoff such that atoms further away than $1.2$ times the nearest neighbor distance from the absorbing site were not considered. This cutoff was chosen as a balance between prediction accuracy and computational cost. The CSM was applied to each absorbing site, and the polyhedron with the lowest CSM value was chosen as the site LCE label.

We restricted the LCE labels to only tetrahedral (T4), square pyramidal (S5) and
octahedral (O6) geometries, because across the eight transition metal families these are the most abundant LCEs, often by an order of magnitude more than the rest. The class breakdown of
the dataset is presented in Table~\ref{tab:nsites}. The total number of site-specific spectra is on average
a few thousand per atom type, with V (3366) and Mn (3493) the most abundant and Cu (839) the least abundant. It should be noted that one can expand the dataset by adding new structures from other material databases, generating artificial structures or introducing additional class labels.
Furthermore, amongst all three chosen classes, O6 dominates, making up about $64$\% of the entire structure database. The impact of the inhomogeneity of the data distribution on the predictive power of the ML model is discussed in Section~\ref{resdisc}.

\begin{table}[!htb]
\caption{\label{tab:nsites}%
The distribution of classes in the structure database used for training the machine learning models.
}
\begin{ruledtabular}
\begin{tabular}{lcccr}
\textrm{Absorber}&
\textrm{T4}&
\textrm{S5}&
\textrm{O6}&
\textrm{Total}\\
\colrule
\text{Ti} & 271 & 359 &  1562& 2192\\
\text{V} & 948 & 412 &  2006& 3366\\
\text{Cr} & 396 & 121 &  902& 1419\\
\text{Mn} & 502& 657 & 2334 & 3493\\
\text{Fe} & 797 & 319 & 1874 & 2990\\
\text{Co} & 583 & 227 & 1428 & 2238\\
\text{Ni} & 246 & 163 &  1238& 1647\\
\text{Cu} & 290 & 183 &  366& 839\\
\hline
\text{Total} & 4033 & 2441 & 11710 & 18184
\end{tabular}
\end{ruledtabular}
\end{table}

\subsection{Training Machine Learning Models \label{training machine learning models}}
The core machine learning algorithm (see Figure~\ref{fig:QAQC})
consists of an optional 1-D convolutional layer followed by
three fully connected, feed-forward hidden layers with 90, 60 and 20 neurons, ending with a
softmax output layer of 3 neurons. The input layer of the neural network is the XANES spectrum scaled to zero mean and unit variance on
a standardized grid of 100 entries, and the output determines the target vector, which contains the
probabilities of the three LCE classes (T4, S5 and O6) computed from the softmax function.
All neurons use the rectified linear unit (ReLU) activation function and a $30$\% dropout to
guard against over-fitting. ML models with and without the 1-D convolutional layer
are referred to as the convolutional neural network (CNN) and multi-layer perceptron (MLP),
respectively.
The optional convolutional layer
contains 8 filters and a kernel (sliding window) size of 10, stride of 1 and max-pooling size
of 2, and takes as input spectral data processed in an identical manner to that of the MLP.
CNN's inherently assume correlations between nearby data points, and being a down-sampling and
pooling technique, sacrifice resolution in favor of invariance to the precise location of input
data.
The algorithm determines trained parameters by minimizing a categorical cross-entropy loss
function using the Adam optimizer \cite{Diederik2014}.
Mini-batch sizes of 32 were used during 50 full passes
(epochs) of the training data. All training and evaluations were performed using Keras~\cite{Chollet2015} with a TensorFlow~\cite{Martin2015} backend.

For each absorbing site, we used statistical boostrapping. $90$\% of the database was
used for training ML models and the remainder for testing. These subsets
were selected randomly in a stratified manner:
the proportion of each class in both the training and testing sets was
always the same. In order
to generate a statistical estimate on the accuracy of the classifier, we
sampled the testing data with
replacement over $10$ folds and report the averaged results in Figure~\ref{fig:prime_res}.
To make full
use of all available information, data included once in a testing set, were not used in any
future testing sets. We found that testing results are mostly invariant to the chosen neural
network architecture assuming enough training parameters were included. Therefore, the fixed 3-layer MLP
with an optional convolutional layer (and associated hyperparameters) was used throughout all experiments.

\section{Results and Discussion \label{resdisc}}
In the following, we present our LCE classification study through visual
inspection, principal component analysis (PCA)~\cite{Pearson1901}
and analysis of the machine learning classifiers (MLCs). We
demonstrate that MLCs can accurately predict LCE classes from synthetic
XANES data generated by FEFF. We further discuss the relevance of this study
based on synthetic data to the real challenge of the LCE classification of
experimentally measured XANES spectra.

\subsection{Visual Inspection of the Spectral Database}
\begin{figure*}[!htb]
\includegraphics[scale=0.6]{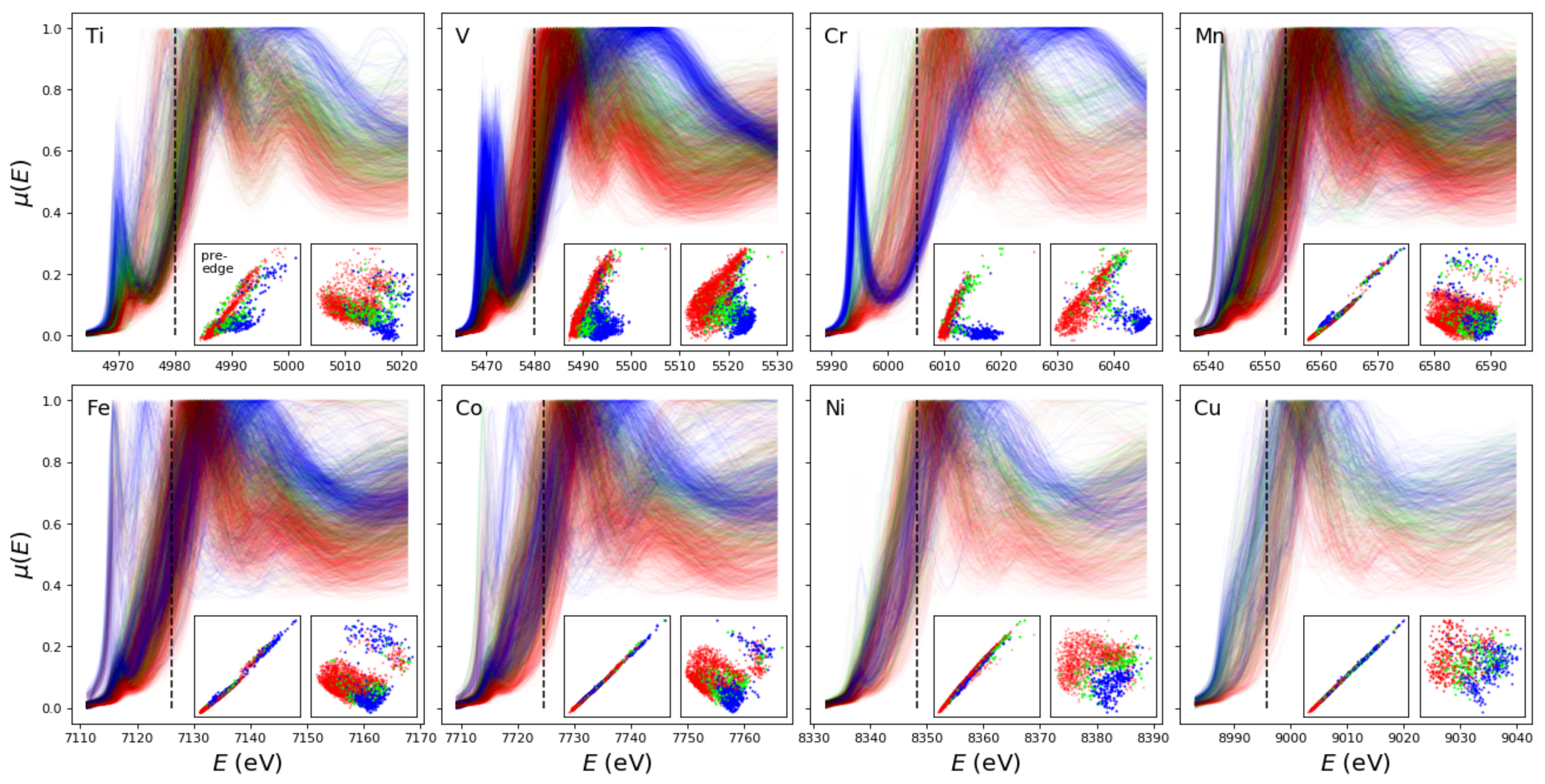}
\caption{\label{fig:all_dat} 
FEFF K-edge XANES database of eight 3$d$ transition metal families. Spectra were spline-interpolated
onto discretized grids of 100
points, and scaled on the vertical axis such that the maximum value is 1 (prior to shifting the mean to 0 and scaling to unit variance). The intensity of the colors scales
inversely with the number of entries per class to aid visualization. 
Principal component analysis of two regions is shown in the two insets: the full feature space (lower right),
and only the pre-edge region (lower center, discussed in Section~\ref{resdisc}). The $x$ and $y$-axes correspond to the first and second principal axes.
The cutoff for the pre-edge region is
delineated by the vertical dashed line. Classes are color coded as follows: blue for tetrahedral (T4), green for 
square pyramidal (S5) and red for octahedral (O6).}
\end{figure*}
The FEFF K-edge XANES database of eight $3d$ transition metal elements (from Ti to Cu)
is shown in Figure~\ref{fig:all_dat}, color coded by LCE class (T4: blue; S5: green; O6: red).
There are noticeable trends in the raw spectra that can be detected by visual inspection, prior to a more in-depth analysis. 
Overall, early $3d$ transition metals (e.g. Ti, V, Cr, and Mn) show more intense pre-edge peaks than
late $3d$ transition metals (e.g. Ni and Cu), consistent with the trend from experiment~\cite{Yamamoto2008}. Notably, T4 in Ti, V and Cr oxides exhibit sharp pre-edge peaks at about $4970$,
$5470$, and  $5995$ eV, respectively. The pre-edge peak intensity decreases as the coordination number increases, consistent
with the observations of Farges \emph{et al.}~\cite{Farges1996, Farges1997,Farges2001} and Jackson \emph{et al.}~\cite{Jackson2005}. 
Such qualitative agreement between theory and experiment suggests that spectral analysis of the FEFF database is physically insightful, especially for the LCE classification problem.

In addition to pre-edge features, across the eight elements T4 exhibits the highest \emph{post-edge} intensity, followed by S5 and finally O6. However, the role of post-edge features in LCE
classification has not yet been explored in the literature, which could be an important supplement to
existing pre-edge based methods. We expect that algorithms including a wide energy range in the XANES
spectra can in principle improve the spectral sensitivity to the LCE as compared to those relying solely on pre-edge features.

\subsection{Principal Component Analysis of the Spectral Database}
We further analyzed the spectral database with PCA. Following the standard
notation, $X^k$ is defined as the full set of spectral data for the $k$th
absorbing species with $\mathbf{x}^{jk}$ being the $j$th spectrum in the dataset (a single feature vector
input) after taking zero sample mean and unit variance. Denote
$\mathbf{w}^{1k}$ and $\mathbf{w}^{2k}$ as the first two principal axes in the feature space. We computed
coordinates of spectrum $j$ in the PCA plot, $\mathbf{z}^{jk} = z_1^{jk} \hat x + z_2^{jk} \hat y$, as
\begin{equation} \label{covariance}
    z_\alpha^{jk} =
    \frac{\boldsymbol{x}^{jk} \cdot \mathbf{w}^{\alpha k}}{\max_l{|z_\alpha^{lk}|}},
    \quad \alpha = 1, 2,
\end{equation}
where for clarity the denominator scales $z_\alpha^{jk}$ within $[-1, 1]$.  

To evaluate the significance of the pre-edge features, we truncated the principal axes by applying a cutoff $n_{c}$ to the spectra, such that $x_{n_c}^{jk}$ correspond to the vertical dashed lines in Figure~\ref{fig:all_dat}.
Then PCA was performed for only the pre-edge region along the truncated principal axes, 
\begin{equation} \label{covariance2}
    \tilde{z}_\alpha^{jk} =
    \frac{\sum_{n=1}^{n_c}x^{jk}_n w^{\alpha k}_n}{\max_l{|\tilde{z}_\alpha^{lk}|}},
    \quad \alpha = 1, 2,
\end{equation}
by excluding features beyond $x_{n_c}^{jk}$.
The axes in the plots generated by Eqs.~\ref{covariance} and \ref{covariance2} are scaled in the same
way, so that their clustering patterns can be compared directly. Similar patterns are expected from the full and pre-edge PCA plots, if the pre-edge features dominate the spectral contrast. On the other hand, if the pre-edge features are less significant, there will be weak correlations between two sets of PCA patterns.

Full PCA plots are shown in the lower right insets of Figure~\ref{fig:all_dat}. Overall, a large degree of
clustering is realized, consistent with the observation of distinguishable spectral features from visual inspection. In Ti, V, Cr, Mn, Fe, and Co, most of
the T4 points are located in the lower right corner and O6 points in the upper left corner. The T4 and
O6 points of the Ti, V, and Cr can be easily separated in PCA plots due to their sharp (T4) and negligible (O6) pre-edge features.
In the PCA plots of Mn, Fe and Co, there is also a secondary cluster of T4
points located in the upper right corner. The data distributions in
Ni and Cu are similar to the others, but with a slightly smaller
packing density. In most cases, S5 clusters are intertwined with the other two classes, flanked by T4 on one side and O6 on the other.



The lower center insets in Figure~\ref{fig:all_dat} show the PCA of the pre-edge region.
All classes appear less clustered than the PCA patterns of the
full feature space. While in this case information contained in the feature space has
clearly been reduced, this effect is less prominent in the
early transition metal elements, which still exhibit a large degree of clustering due
to the significant spectral contrast in the pre-edge region.
On the contrary, elements such as Co, Ni and Cu exhibit such severe information loss that 
PCA data points collapse into a linear pattern, which is detrimental to the MLC performance, especially in systems that already exhibit weak spectral contrast.

In summary, visual inspection and PCA suggest that a MLC is likely able to accurately learn the trends in XANES spectra and correlate them to their respective
classes. However, it is unclear whether MLCs can perform equally well for every class in all of the transition metal species we studied.

\subsection{Performance of the Machine Learning Classifiers}
The accuracy of the MLCs for each LCE class is reported using the
$F_1$ score, which is the harmonic mean of the precision $P$ and recall $R,$
\begin{equation} \label{f1}
    F_1 = \frac{2PR}{P + R}, \quad P = \frac{t_+}{t_+ + f_+}, \quad R = \frac{t_+}{t_+ + f_-},
\end{equation}
where $t_+, f_+$ and $f_-$ represent the true positives, false positives and
false negatives in a two-class (2 by 2) confusion matrix.

In order to make a fair assessment of the MLC performance, we need to address the
data imbalance issue in our training set. As seen in Table~\ref{tab:nsites}, the
number of LCEs that conform to the O6 geometry vastly outnumbers the others,
indicating that the accuracy of each class alone might not be the most reliable metric, as it may
be biased due to class imbalances in the training data. We address this problem in two ways: by using the $F_1$ score
instead of the accuracy on a class-by-class basis, and reporting the macro $F_1$ score as a representative metric.

In general, the $F_1$ score is a much stricter metric than the accuracy and
is a better indicator of performance. It accounts for both the precision $P$ (of all predicted positives, cases that
are actually positive) and recall $R$ (out of all the actual positives, cases that are correctly identified) and
dramatically penalizes poor scores in either category (contrary to the mean of the precision and recall). To demonstrate
why this is important, consider the $F_1$ score of the relatively underrepresented S5 class. Suppose that there are
10 S5 and 100 T4 and O6 in the data set, and that the classifier has a 10\% false negative and false positive rate. Accuracy would naturally be 90\%, but this is a terrible representation of the classifier, since 10 non-S5 data points
were predicted as S5, unnaturally inflating the number of predicted positives. On the contrary, the $F_1$ score of 62\%
accounts for this by incorporating the low precision (47\%) into the metric.

In addition to a breakdown by class, the \emph{macro} $F_1$ score ($\overline{F}_1$)
is reported, which is the average of the class-wise $F_1$ scores computed using a one-versus-all approach. The $\overline{F}_1$ score treats each class on equal footing and further penalizes
classifying data in an underrepresented class incorrectly relative to a class with many data points.

\begin{table}[!htb]
\caption{\label{tab:macrof1}%
$\overline{F}_1$ scores for different absorbing species, as the averages of the class-wise $F_1$ scores
(red, green and blue bars) presented in Figure~\ref{fig:prime_res},  both with (CNN) and without (MLP)
the convolutional layer. Comparisons are made between models trained from the full feature space and
reduced feature space corresponding to only the pre-edge region of the spectra.
}
\begin{ruledtabular}
\begin{tabular}{lcc|cc}
& \multicolumn{2}{l}{Full Feature Space} & \multicolumn{2}{c}{Pre-edge Only}\\
\hline
\textrm{Element}&
\textrm{MLP}&
\textrm{CNN}& \textrm{pre-MLP}&\textrm{pre-CNN}\\
\colrule
\text{Ti} & 0.83(2) & 0.84(2) & 0.73(4) & 0.78(3) \\
\text{V} & 0.86(1) & 0.86(2) & 0.77(2) & 0.79(3) \\
\text{Cr} & 0.87(2) & 0.87(3) & 0.72(4) & 0.75(4) \\
\text{Mn} & 0.83(2) & 0.85(2) & 0.62(4) & 0.68(3) \\
\text{Fe} & 0.85(2) & 0.86(3) & 0.57(2) & 0.63(3) \\
\text{Co} & 0.85(3) & 0.87(2) & 0.59(3) & 0.64(3) \\
\text{Ni} & 0.87(1) & 0.88(3) & 0.61(5) & 0.66(4) \\
\text{Cu} & 0.86(2) & 0.86(2) & 0.50(4) & 0.64(7)\\
\colrule
\text{Average} & 0.85(1) & 0.86(1) & 0.64(1) & 0.70(1)
\end{tabular}
\end{ruledtabular}
\end{table}

\begin{figure*}[!htb]
\includegraphics[scale=0.601]{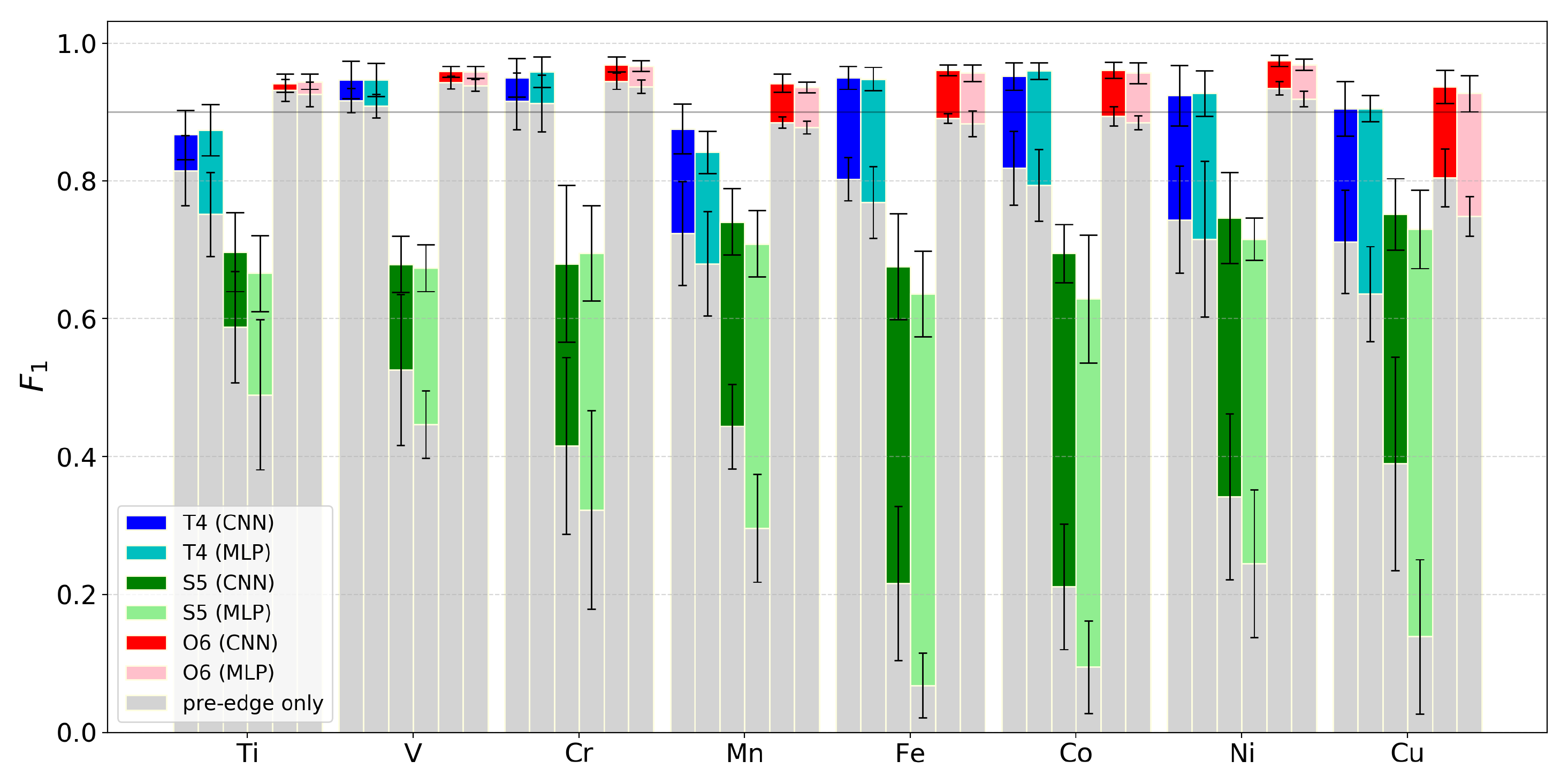}
\caption{\label{fig:prime_res} 
Class-wise $F_1$ scores calculated using different machine learning models
(CNN/MLP) for T4 (blue/cyan), S5 (green/light green) and O6 (red/pink) local coordination environments in 
eight $3d$ transition metal elements. While the full height of each bar represents the results trained on the full feature space, gray bars overlaid on top with lower $F_1$ values represent the results for the models trained only on the pre-edge region. For example, for the Co S5 CNN, the $F_1$ score reported for training on the full
spectral space is about 0.7, but decreases sharply to about 0.2 when trained on the pre-edge region only. Error bars correspond to
standard deviation; the ones with wider caps correspond to the full feature space.}
\end{figure*}

As clearly shown from the $\overline{F}_1$ scores in Table~\ref{tab:macrof1},
MLCs can classify the LCEs of all eight $3d$ transition metal families very accurately. Uncertainties reported in the last digit of Table~\ref{tab:macrof1} and error bars in Figure~\ref{fig:prime_res} 
correspond to the standard deviation calculated from ten different trained models. CNNs and MLPs perform equally well, with very close $\overline{F}_1$ scores of 0.86 and 0.85, respectively. 
The class-wise $F_1$ scores
are plotted in Figure~\ref{fig:prime_res}. Notably, MLCs can reach over $90$\% accuracy
on the T4 (CNN: 0.92; MLP: 0.92)  and O6 classes (CNN: 0.96; MLP: 0.95), which can be understood from the observation of raw spectra. The strong pre-edge peak intensity is a
signature of the T4 configuration in, e.g., Ti, V and Cr. Conversely, the lack of a
significant pre-edge peak is a clear indicator of an O6
configuration. In these cases, it is likely that the pre-edge features are sufficient to
distinguish O6 from T4. On the contrary, the spectral contrast between T4 and O6 is
very low in the pre-edge region in late transition metal elements, especially in Ni
and Cu. It is remarkable that MLCs can achieve the same accuracy for T4 and O6 in
late transition metal elements. Such a universally good performance underscores the
ability of the MLCs to extract spectral descriptors without human bias in the full energy range, including the pre-, main- and post-edge regions. Moreover, the relatively small overall error margin is a testament to the reliability and robustness of the classifier across many trained models.

Relatively speaking, the S5 classification is less successful than T4 and O6, with an
overall accuracy of $\sim 0.70$ (CNN: 0.71; MLP: 0.68), as shown in
Figure~\ref{fig:prime_res}. The weaker performance of MLCs on the S5 class can be
explained by Figure~\ref{fig:all_dat}, where data associated with the S5 class lay between
those in T4 and O6 in both the spectral and principal component space, making them
more difficult to identify.

\subsection{Importance of Features Beyond the Pre-edge}
The ability of MLCs to accurately classify late transition metal
oxides that lack prominent pre-edge features suggests that 
features beyond the pre-edge region play an important role in the
neural network model. This hypothesis is supported by the PCA results
shown in the insets of Figure~\ref{fig:all_dat}. 
In the late transition metals, while the full
spectra can be effectively clustered in two-component PCA, the same analysis of pre-edge spectra displays a completely different linear pattern resulting from substantial information loss.

To gain further insight, we train MLCs with identical architectures using \emph{only} the pre-edge
region defined by energies below the dashed lines in Figure~\ref{fig:all_dat}, which we refer to as the pre-MLCs (pre-CNNs and pre-MLPs). 
In principle, if $\overline{F}_1$ scores of the pre-MLCs are close to those trained
on the full spectra, then the pre-edge features are sufficient to classify the LCE for that absorbing element. Conversely,
a significant drop in the $\overline{F}_1$ scores of the pre-MLCs would be a clear indication that features beyond the pre-edge region play a significant role in the MLCs.

As shown in Table~\ref{tab:macrof1}, the average $\overline{F}_1$ score in pre-MLCs drops significantly by about $20$\%
(from $0.86$ to $0.70$ in CNN and from $0.85$ to $0.64$ in MLP), as compared to
MLCs trained on the full spectral space. The class-wise $F_1$ scores of pre-MLCs which are consistently lower than that of regular MLCs are shown in Figure~\ref{fig:prime_res} as the
gray bars. We quantify this accuracy degradation by
$$\Delta = F_1(\textrm{MLC})-F_1(\textrm{pre-MLC})$$ 
and summarize the results averaged
over early (Ti, V, Cr, and Mn) and late transition metal elements (Fe,
Co, Ni and Cu) in Table~\ref{tab:deltas}. First,
$\Delta$ in late transition metals is more than doubled compared to early
transition metals. Second, among the three
classes, $\Delta$ of O6 is the smallest ($< 0.10$). It increases significantly for T4 in late transition metals to $0.16$ in the pre-CNN ($0.21$ in the pre-MLP) and finally reaches the largest values for S5, at $0.22$ ($ 0.30$) in early transition metals and $0.44$ ($0.57$) in
late transition metals. The results in
Figure~\ref{fig:prime_res} and Tables~\ref{tab:macrof1}-\ref{tab:deltas} highlight 
the  critical importance of features beyond the pre-edge region in accurately classifying LCEs, especially for late transition metals. The effects are the largest in the S5 class,
which is rather characterless in the pre-edge region, showing neither very strong (like T4) nor very weak (like O6) pre-edge intensities.

\begin{table}[!htb]
\caption{\label{tab:deltas}%
The class-wise difference between the $F_1$ score evaluated over the entire
feature space and over only the pre-edge region
($\Delta$). Results are averaged over the early
(Ti, V, Cr and Mn) and late (Fe, Co, Ni and Cu) transition metal elements.}
\begin{ruledtabular}
\begin{tabular}{lccc|ccc}
& \multicolumn{3}{c}{pre-MLP} & \multicolumn{3}{c}{pre-CNN}\\
\hline
&\textrm{T4}&\textrm{S5}&\textrm{O6}&\textrm{T4}&\textrm{S5}&\textrm{O6}\\
\colrule
\text{Early} & 0.09(3) & 0.30(5) & 0.03(1) & 0.07(3) & 0.22(5) & 0.03(1)\\
\text{Late} & 0.21(4)  & 0.57(5) &0.10(1) & 0.16(3) & 0.44(7) &0.08(1) 
\end{tabular}
\end{ruledtabular}
\end{table}

We note that unlike the case of regular
MLCs, the pre-edge CNN averaged over all absorbing species ($\overline{F}_1=0.70$) outperforms the corresponding pre-edge MLP ($\overline{F}_1=0.64$) substantially.
In the most extreme situation of S5, the pre-CNN ($F_1=0.48$) is $23$\% more accurate than the pre-MLP ($F_1=0.39$) in early transition metals, and it is more than doubled in late transition metals with $F_1=0.28$ $(0.12)$ for the pre-CNN (pre-MLP).
The substantially better performance of the pre-CNN is likely caused by
the use of the convolutional filter, which makes the CNN able to learn
subtle pre-edge features from augmented data more effectively than the
MLP.

\subsection{Discussion}
The MLCs described so far were trained on computational FEFF XANES spectra.
Developing MLCs that can classify the LCE of a broad range of material families using experimental XANES
spectra is a more challenging task that is beyond the scope of the current work. Nonetheless, in this
section we discuss several key issues that need to be addressed
in order to achieve this goal, including validation of the theory, edge alignment
of the simulated spectra, and the variations in the spectral intensity.

\begin{figure}[!htb]
\includegraphics[scale=0.395]{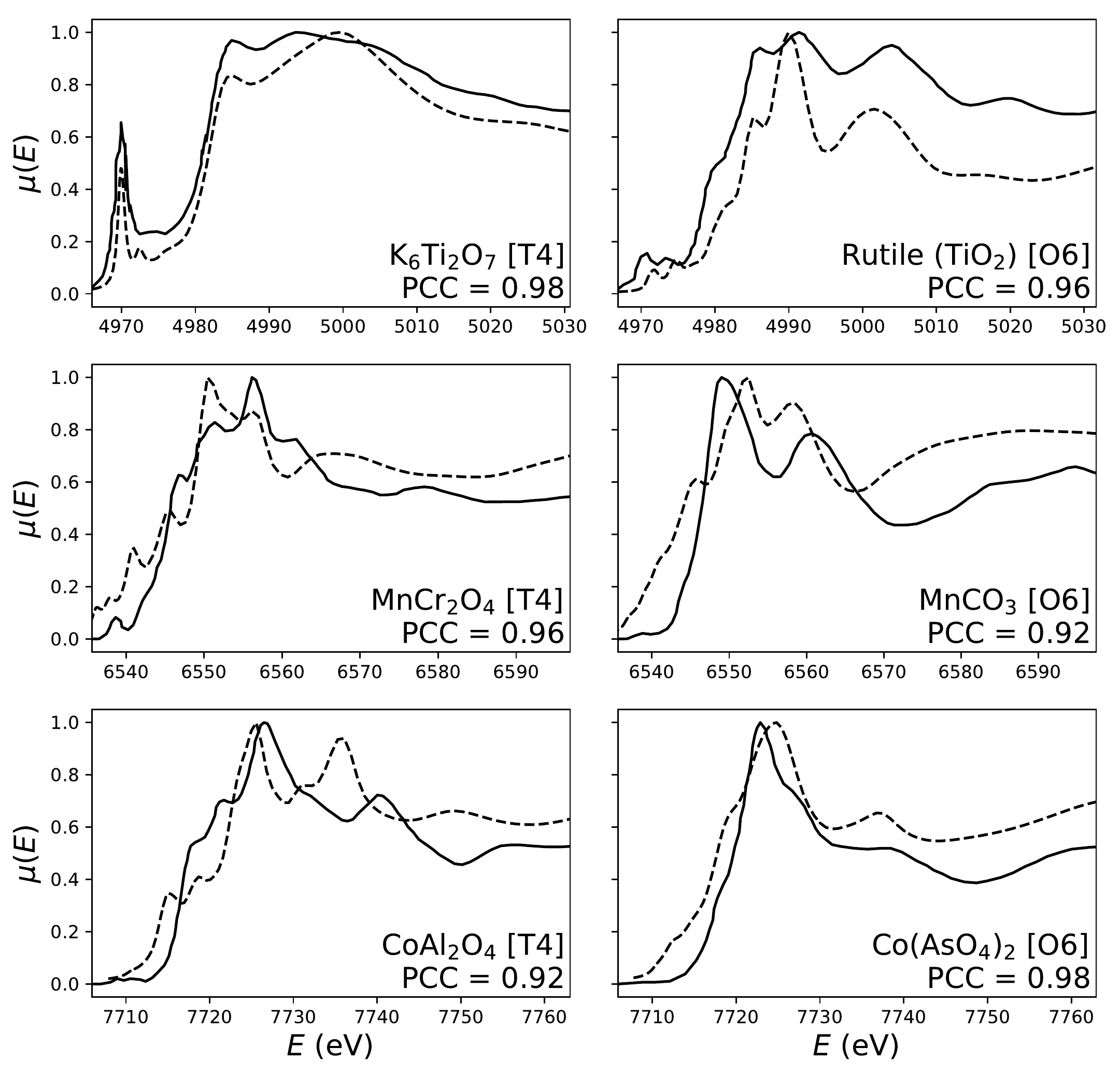}
\caption{\label{fig:exp_comparison} Comparison between representative experimental
(solid) and FEFF (dashed) XANES spectra. FEFF calculations are
shifted on the energy axis to maximize such that the Pearson correlation coefficient (PCC) in order to find the best match between experimental and theoretical XANES. The experimental spectra of K$_6$Ti$_2$O$_7$ and
rutile were extracted from Farges \emph{et al.}~\cite{Farges1997}, and the experimental spectra of both pairs of Mn and Co oxides from Manceau \emph{et al.}~\cite{Manceau1992}.}
\end{figure}

In order to appy MLCs trained on synthetic data to experimental
spectra, it is very important to validate the theory such that
the computational spectra can faithfully reproduce experimental spectral
features. To this end, we compare FEFF spectra with experimental spectra 
on a small number of oxides: K$_6$Ti$_2$O$_7$, rutile (TiO$_2$), MnCr$_2$O$_4$, MnCO$_3$, CoAlO$_4$ and
Co(AsO$_4$)$_2$ in Figure~\ref{fig:exp_comparison}. 
This list is not meant to be exhaustive.
Within this small sample, while the overall shape and major peaks are
well reproduced by FEFF, there
are noticeable differences in the spectral details, including the peak positions
and relative intensities of different peaks. Furthermore, the degree of agreement is system-dependent.
As shown in Figure~\ref{fig:exp_comparison}, the optimal Pearson correlation coefficients (PCCs) between FEFF and experiment range from $0.92$ to $0.98$. 

Despite the relatively high PCC scores,
MLCs trained on spectra at the FEFF level of theory as such cannot reliably classify
experimental spectra.
It is necessary to generate the computational XANES database with more accurate methods and conduct a systematic benchmark of theory against experiment. However, the
computational expense of these calculations grows quickly with the complexity of
the methods, which could in practice limit the level of theory used to generate the training
set. A good compromise would involve developing robust ML algorithms for a
physically sound but numerically imperfect training set. It may also be possible to augment
the computational spectral database with a subset of experimental data
and apply ML techniques that can handle a hybrid database, such as transfer learning.

\begin{figure}[!hbt]
\includegraphics[scale=0.395]{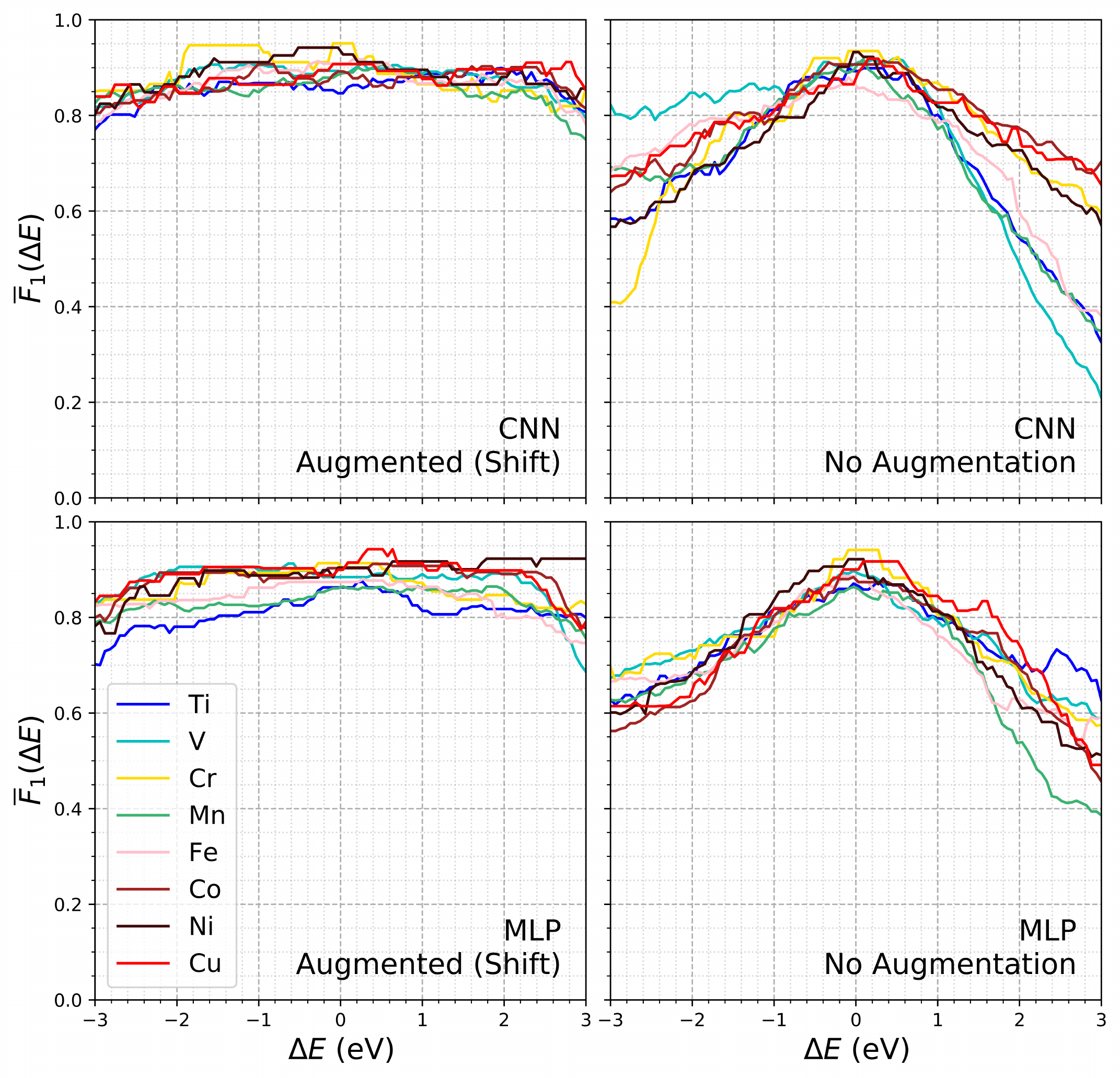}
\caption{\label{fig:shift_1_plot} $\overline{F}_1$ score as a function of 
the amount of energy shift ($\Delta E$) applied to the test set. The two graphs on the left
illustrate the results of MLCs trained on augmented data (as presented thus far: $\pm 1$ and $2$
eV, generating 5 times the base amount of training data), while those on the right illustrate
MLCs trained without augmenting the training set.}
\end{figure}

Another open question in computational XANES is the edge alignment of computational spectra with
experimental spectra, because current first-principles electronic structure methods have difficulty predicting accurate absolute onset energies.
This issue stems from the use of pseudopotentials and/or approximations to the electron
self-energy and core-hole final state effects. Therefore, XANES calculations are often
analyzed with the relative energy scale or after they are manually aligned with
reference experimental spectra. However, the energy shift in the spectral alignment with
respect to reference experimental spectra could be system-dependent, which warrants further
study.

In order to investigate the impact of the edge alignment on the performance of MLCs, we shift the test set by up to $\pm 3$ eV to study the transferability of MLCs against the shifted data. We note that a shift of $3$ eV in energy is quite significant, as the energy range of the pre-edge region is about $10$ to $15$ eV. As shown in Figure~\ref{fig:shift_1_plot}, the MLCs are very robust against the energy shift, as the $\overline{F}_1(\Delta E)$ curves are almost flat. The CNN slightly outperforms MLP with $\overline{F}_1$(CNN)$> 0.8$ in most of the range of $\Delta E$ for all eight elements. The robustness of the MLCs results from the data augmentation we applied in the training set with energy shifts of $\pm 1$ and $\pm 2$ eV as described in the Section~\ref{methods}. If we apply the same test on MLCs developed from the base training set without data augmentation, the accuracy deteriorates quickly after $|\Delta E| > 1$ eV, as shown in Figure~\ref{fig:shift_1_plot}.

\begin{figure}[!htb]
\includegraphics[scale=0.395]{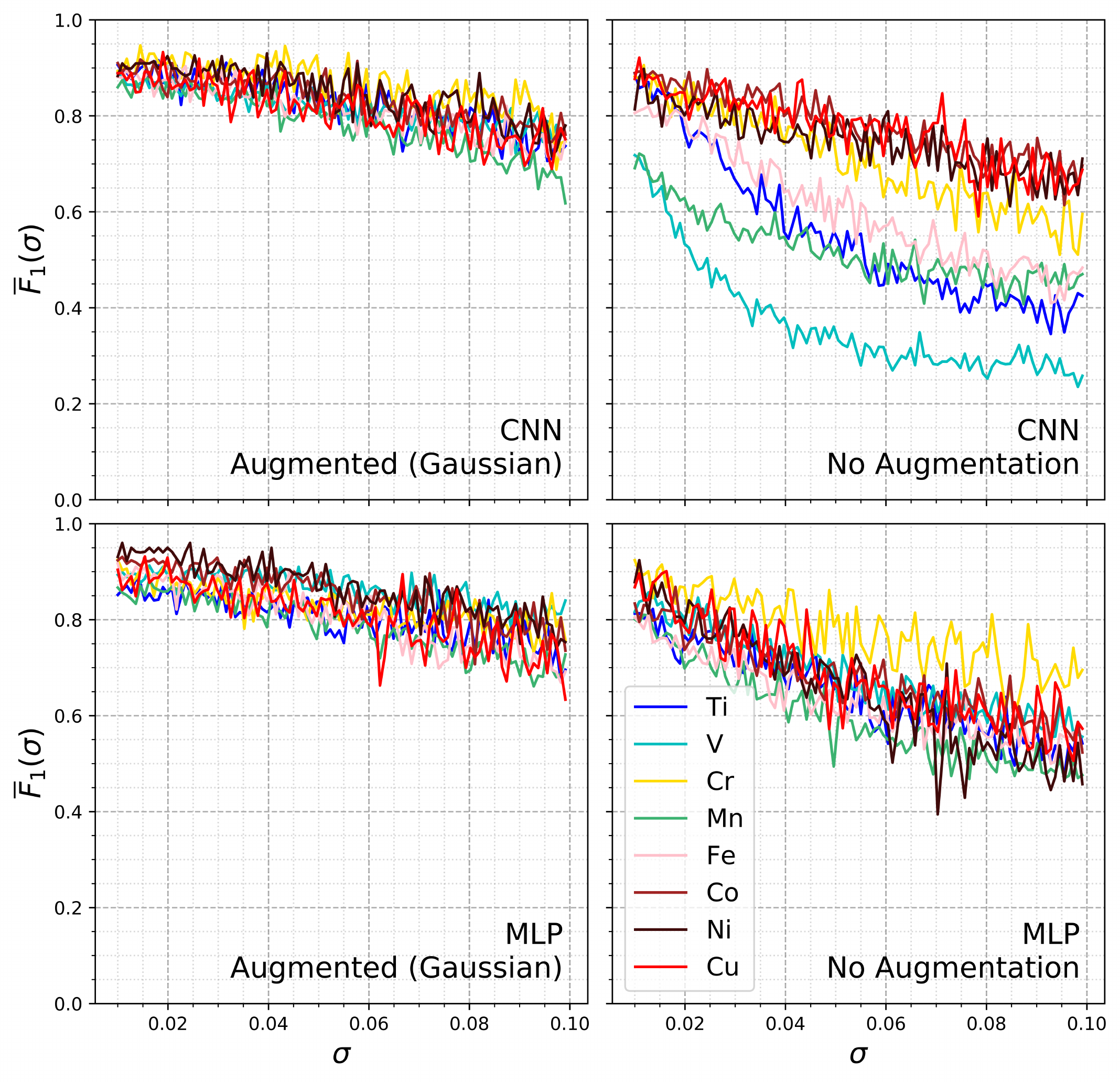}
\caption{\label{fig:shift_2_plot} $\overline{F}_1$ score as a function of the
standard deviation ($\sigma$) used to generate Gaussian random noise, introduced
into every point $E$ in the spectra $\mu(E)$. The two graphs on the left illustrate
the results of MLCs trained on a training set augmented with Gaussian random noise of
$\sigma  = 0.03$ (also generating 5 times the base amount of training data), and those on the
right illustrate MLCs trained without data augmentation.}
\end{figure}
Additionally, we have seen that the spectral intensity of theoretical XANES spectra may not match perfectly with experiments. On top that, the intensity of experimental XANES spectra is subject to several uncertainties from sample preparation and various instrumental factors, such as type and mosaic
spread of the monochromator crystals, source sizes, slit heights and beam instabilities~\cite{Farges1997} and the resolution of the apparatus~\cite{Yoshitake2003}. Therefore, experimental spectra of the same materials that are measured using different samples or collected at different beamline settings may have
slightly different intensity profiles. 
To investigate the impact of the uncertainties in the spectral intensity on
the MLCs, we introduced Gaussian random noise with standard deviation $\sigma$ to the spectral intensity centered around $\mu(E)$ for every $E$ on the energy grid. 
To isolate the effects of the Gaussian random noise, we test the MLCs trained without augmentation accounting for the energy shift. As shown in Figure~\ref{fig:shift_2_plot}, the overall $\overline{F}_1$ score decays quickly
with the increasing $\sigma$ in both the CNN and MLP, with V trained using the CNN suffering the most. After we augmented the training set with spectra containing the Gaussian random noise with $\sigma=0.03$, the $\overline{F}_1$ score decays much more slowly with increasing $\sigma$.

From the analysis above, one can clearly see that MLCs perform dramatically better when the training set is augmented, which is a sensible result. It is interesting to note that the CNN underperforms relative to the MLP without data
augmentation, specifically for early transition metals that exhibit strong pre-edge peaks. For example, the $\overline{F}_1$ score of the unaugmented CNN for Ti and V drops to $\sim 0.45$ and $0.25$, respectively, at $\sigma=0.1$. This trend is not entirely counterintuitive for two reasons. First, the CNN as described in Subsection~\ref{training machine learning models} is not completely shift-invariant, and without proper data augmentation it may not learn how to account for small perturbations. Second, as shown in Table~\ref{tab:deltas}, the CNN relies more on the pre-edge region than the MLP. Since shifting the location of the pre-edge peak strongly affects the pre-edge spectral features, a sizable drop in performance is to be expected. A similar argument may be made for the effects of Gaussian random noise, which can artificially distort both the shape and location of peaks.

\section{Conclusion}
We propose a new computational framework to perform element-specific classification of local chemical environments from XANES spectra. In addition to the construction of structure and spectral databases and structural labels, a central element of this framework is unraveling the correlation between spectral features and local chemical environments systematically
using machine learning classifiers. As proof-of-principle, we applied our method to the computational XANES database of eight 3$d$ transition 
metal elements generated by the FEFF code and achieved a high average macro $F_1$ score of $0.86$. 
Our method can reliably capture not only the prominent pre-edge features, but also the less characteristic spectral features beyond the pre-edge region. 
We showed that features beyond the pre-edge region turn out to be very important to the accuracy of the classification,
especially for late transition metal elements. The ability to extract key structural information in the full spectral range makes our machine learning-based method
more robust and transferable than empirical fingerprint methods based solely on the pre-edge region. As an important starting point,
our work will motivate future research on the problem of classification of local chemical environments on experimental measured spectra.

\begin{acknowledgments}
This research used resources of the Center for Functional Nanomaterials, which is a U.S. DOE Office of Science Facility, and the Scientific Data and Computing Center, a component of the BNL Computational Science Initiative, at Brookhaven National Laboratory under Contract No. DE-SC0012704. This research is also, in part, supported by Brookhaven National Laboratory LDRD (Lab Directed Research and Development) 16-039. M.R.C. acknowledges support from the U.S. Department of Energy through the Computational Sciences Graduate Fellowship (DOE CSGF) under grant number: DE-FG02-97ER25308. The authors acknowledge fruitful discussions with Mark Hybertsen, Anatoly Frenkel, Bruce Ravel, Klaus Attenkofer, Eli Stavitski, Xiaochuan Ge, Sencer Selcuk and Marco Baity-Jesi.
\end{acknowledgments}

\newpage
\newpage
\providecommand{\noopsort}[1]{}\providecommand{\singleletter}[1]{#1}%

\end{document}